\documentclass[11pt, oneside]{article}   	
\usepackage{geometry}                		
\usepackage{graphicx}				
\usepackage{amssymb}
\usepackage{amsmath}
\usepackage{subfig}
\usepackage{caption}
\usepackage{comment}

\usepackage[utf8]{inputenc}
\usepackage{amsmath}
\numberwithin{equation}{section}
\begin{document}
\title{Pandemic modeling and the renormalization group equations: \\Effect of contact matrices, fixed points \\and nonspecific vaccine waning}
\author{ Michael McGuigan \\
Brookhaven National Laboratory}
\date{}
\maketitle
\begin{abstract}
In this paper we find common features between the equations that are used for pandemic or epidemic modeling and the renormalization group equations that are used in high energy physics. Some of these features include the relation of contact matrices in pandemic modeling and operator mixing in the renormalization group equations. Another common feature are the use of flow diagrams and the study of fixed points both in pandemic modeling and in evolution under  renormalization group equations. We illustrate these relations through the study of some cases of interest to the current COVID-19 pandemic. These include pandemic modeling with mixing between different age groups and also contact matrices associated with contact between countries. For the final example we study the effect on mortality of waning from nonspecific vaccines which are designed to combat different pathogens but nevertheless may lessen the severity and mortality of COVID-19 infections.


\end{abstract}
\newpage

\section{Introduction}

It is well known that in physics phenomena that are governed by the same type of equations can benefit from the sharing of techniques for the solution and visualization of the solutions to those equations. For example the simple harmonic oscillator and electronic circuits are described by similar equations and insight from one equation can be applied to the other. The first order nature of the differential equations used in compartment models for epidemic modeling draw a striking resemblance to the renormalization group  equations used to describe high energy physics models. Essentially each compartment or the number of  class of individuals in the model correspond to a coupling constant in the high energy physics model. One can even go beyond this and draw a closer relation to scaling structure of virus spreading and evolution of a physics model under change in scale \cite{Tarpin}\cite{DellaMorte:2020wlc}. In this paper we illustrate the relation between. the techniques used in high energy physics and apply these to cases of interest to pandemics with similar characteristics to the current COVID-19 pandemic. 

This paper is organized as follows. In section one we give a basic introduction to our approach. In section two we give the basic equations that are used in pandemic and epidemic modeling based on classifying individuals in different compartments. In section three  we give the basic renormalization group equations of high energy physics associated with coupling constants. We show how basic solution techniques and visualizations involving flow diagrams can also be applied to compartment models of pandemic modeling and how these can be used to identify fixed points of final state equilibrium in pandemic modeling. We also draw a parallel between the  between pandemic contact matrices and operator mixing where a coefficient of an operator that may be zero at some scale nevertheless can become nonzero at a different scale through mixing with other operators in the theory. In section four we illustrate the above relations using contact matrices between different age groups using demographics and pandemic characteristics similar to the current COVID-19 pandemic. In section five we further illustrate the techniques using contact matrices between different countries which encounter the virus through delayed interaction and visualizing the curves for infection and mortality during the pandemic. In section six we study the effect of nonspecific vaccines that do not prevent infection but may lessen the severity and mortality of the pandemic by studying the model with separate age groups both with and without the use of a nonspecific vaccine. Finally in section seven we state the main conclusions of the paper and opportunities for further exploration of the common features of pandemic modeling and remormalization group equations.

\newpage

\section{Equations of compartment models of epidemics}

Starting with the pioneering work of Kermack and McKendrick \cite{Kermack} epidemic models can be described by nonlinear differential equations. For example the basic equations of the compartmental $ SIR$ model of epidemics are as follows \cite{Arino1}\cite{Arino2}\cite{Brauer}\cite{Weiss}:
\begin{equation}\begin{aligned}
& S'(t) =  - \beta S(t)I(t) \\
&I'(t) = \beta S(t)I(t) - \gamma I(t)\\
& R'(t) = \gamma I(t)\\
\end{aligned}\label{eqn2.qo}
\end{equation}
with initial conditions
\begin{equation}S(0) = {S_0},I(0) = {I_0},R(0) = {R_0}\end{equation}
Here $S$ is the fraction of the population of susceptible individuals, $I$ is the fraction of infected individuals and $R$ is the fraction of recovered individuals. In these equations $\beta$ and $\gamma$ are constants associated with the epidemics that determine the likelihood of infection and the rate of recovery.

Another model that we will use is the $SIRD$ model which is similar to the $SIR$ model but takes into account mortality associated with the infection. For the $ SIRD $ model the compartment equations are:
\begin{equation}\begin{aligned}
& S'(t) =  - \beta S(t)I(t) \\
&I'(t) = \beta S(t)I(t) - \gamma I(t)- \mu I(t)\\
& R'(t) = \gamma I(t)\\
& D'(t) = \mu I(t)\\
\end{aligned}\label{eqn2.qo}
\end{equation}
with initial conditions
\begin{equation}S(0) = {S_0},I(0) = {I_0},R(0) = {R_0}, D(0) = D_0\end{equation}
Here $S$ is the fraction of the population of susceptible individuals, $I$ is the fraction of infected individuals, $R$ is the fraction of recovered individuals and $D$ is the fraction of deceased individuals. In these equations $\beta$, $\gamma$ and $ \mu$ are constants associated with the epidemics that determine the likelihood of infection, the rate of recovery and the rate of dying from the infection.

For small number of compartments these epidemic equations are not difficult to solve numerically using software such as Mathematica or MATLAB \cite{Knipl}. In figure 1 we show the numerical solution for an epidemic described by an $SIR$ model for parameters $ ( \beta = 3/2, \gamma = 2/3 )$ with initial conditions $(S_0 = .999, I_0 = .001, R_0 =0)$.

\begin{figure}
\centering
  \includegraphics[width =  .75\linewidth]{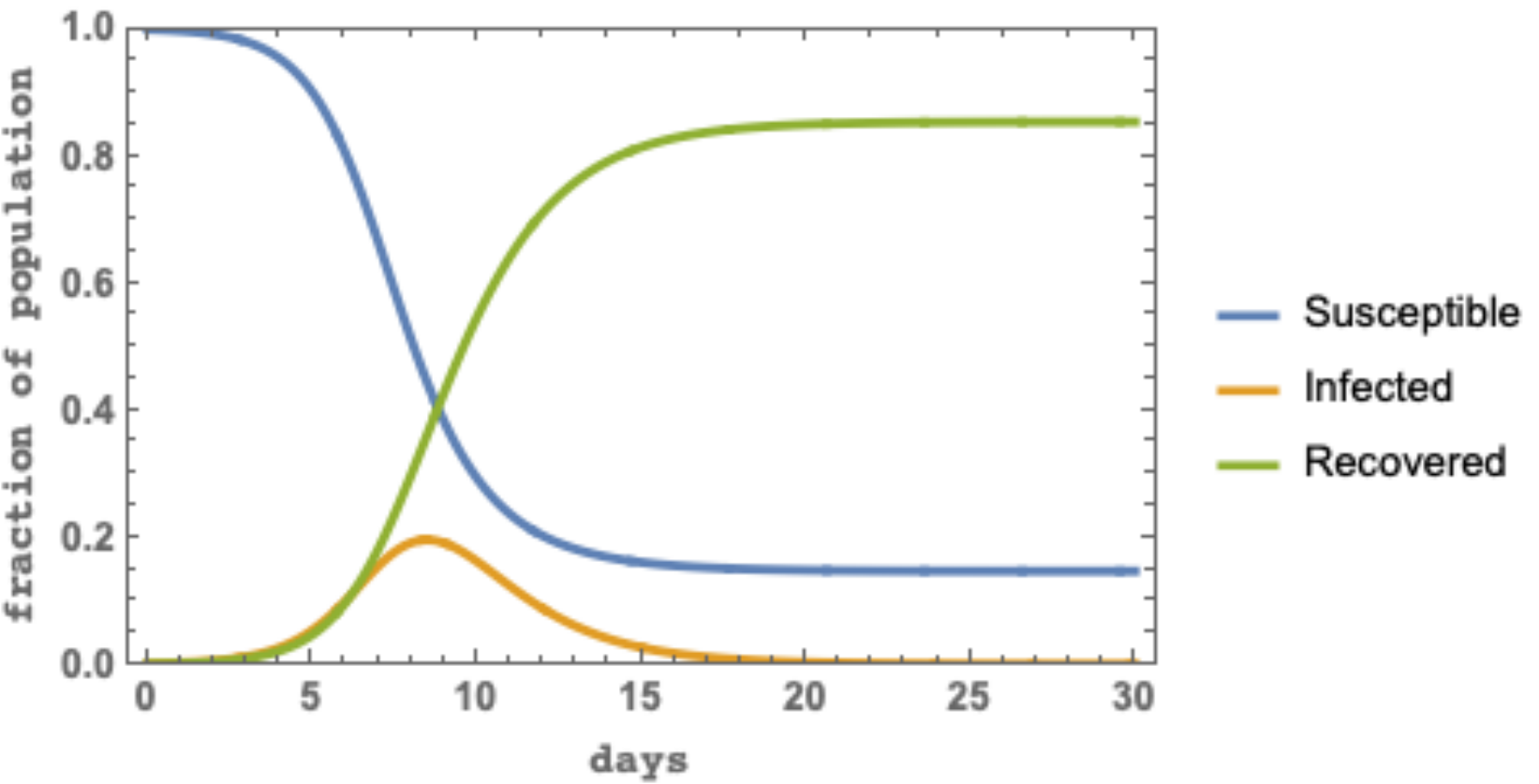}
  \caption{Numerical solution for an epidemic described by an $SIR$ model for parameters $ ( \beta = 3/2, \gamma = 2/3 )$ with initial conditions $(S_0 = .999, I_0 = .001, R_0 =0)$.}
  \label{fig:Radion Potential}
\end{figure}

\section{Equations of the renormalization group for high energy physics}

In the compartmental models of epidemics one has different compartments associated with susceptible, infected , recovered and deceased individuals. In the renormalization group equations of high energy physics instead of compartments one has coupling constants that describe interactions in the high energy physics model. Instead of evolving the system of equations in time as in the equations of epidemic modeling in the renomalization group equation one has evolution with respect to energy scale whose logarithm is denoted as $t$. Perhaps the most important system of renormalization group equations of high energy physics is the equations governing the coupling constants of the standard model of high energy physics \cite{Jegerlehner:2014xxa}. These can be written as:
\begin{equation}\begin{aligned}
&\frac{{d\lambda }}{{dt}} = {\beta _\lambda }\\
&\frac{{dy_t}}{{dt}} = {\beta _{y_t}}\\
&\frac{{dg_1 }}{{dt}} = {\beta _{g_1} }\\
&\frac{{dg_2 }}{{dt}} = {\beta _{g_2} }\\
&\frac{{dg_3 }}{{dt}} = {\beta _{g_3} }\\
\end{aligned}\label{eqn2.qo}
\end{equation}
with the right hand side give by the beta functions of the theory whose coefficients are computed using Feynman diagrams and are given by:
\begin{equation}\begin{aligned}
&{\beta _\lambda } = \frac{1}{{{{\left( {4\pi } \right)}^2}}}(\lambda (24\lambda  + 12y_t^2 - 9g_2^2 - 3g_1^2) - \left( {6y_t^4 - \frac{9}{8}g_2^4 - \frac{3}{8}g_1^4 - \frac{3}{4}g_1^2g_2^2} \right))\\
&{\beta _{{y_t}}} = \frac{1}{{{{\left( {4\pi } \right)}^2}}}({y_t}(\frac{9}{2}y_t^2 - 8g_3^2 - \frac{9}{4}g_2^2 - \frac{{17}}{{12}}g_1^2))\\
&{\beta _1} = \frac{1}{{{{\left( {4\pi } \right)}^2}}}(\frac{{41}}{6}g_1^3)\\
&{\beta _2} =  \frac{1}{{{{\left( {4\pi } \right)}^2}}}(- \frac{{19}}{6}g_2^3)\\
&{\beta _3} = \frac{1}{{{{\left( {4\pi } \right)}^2}}} (- 7g_1^3)\\
\end{aligned}\label{eqn2.qo}
\end{equation}
with initial conditions given by:
\begin{equation}\lambda(0)=\lambda_0, y_t(0)={y_t}_0), g_1(0)={g_1}_0, g_2(0)={g_2}_0, g_3(0) = {g_3}_0\end{equation}
In these equations $\lambda$ is the coupling constant of the Higgs boson, $y_t$ is the coupling constant of the top quark and  $(g_1,g_2,g_3)$ are the coupling constants the electroweak and strong interactions. Looking at the structure of the system of differential equations describing the renormalization group one can see that it describes a system of similar complexity to a five compartment epidemic model. In the following sections we will consider models with eight and twelve compartments of even greater complexity.

One can study a simpler form of renoramlization group equations associated with the Higgs boson coupled to the top quark in a model referred to as the Yukawa model \cite{Molgaard:2014mqa}. For this model one has two coupling constants and the renormalization group equations are given by:



\[\frac{{d\lambda }}{{dt}} = {\beta _\lambda }\]
\begin{equation}\frac{{dy}}{{dt}} = {\beta _y}\end{equation}
with
\begin{equation}\begin{aligned}
&{\beta _\lambda } = \frac{1}{{{{\left( {4\pi } \right)}^2}}}2(12{\lambda ^2} + 2{y^2}\lambda  - {y^4})\\
&{\beta _y} = \frac{1}{{{{\left( {4\pi } \right)}^2}}}\frac{5}{2}{y^3}\\
\end{aligned}\label{eqn2.qo}
\end{equation}
Now defining as in \cite{Molgaard:2014mqa}:
\[{a_\lambda } = \frac{\lambda }{{{{\left( {4\pi } \right)}^2}}}\]
\begin{equation}{a_y} = \frac{{{y^2}}}{{{{\left( {4\pi } \right)}^2}}}\end{equation}
we have:
\[\frac{{d{a_\lambda }}}{{dt}} = {\beta _{{a_\lambda }}}\]
\begin{equation}\frac{{d{a_y}}}{{dt}} = {\beta _{{a_y}}}\end{equation}
with
\begin{equation}\begin{aligned}
&{\beta _{{a_\lambda }}} = 2(12a_\lambda ^2 + 2{a_y}{a_\lambda } - a_y^2)\\
&{\beta _{{a_y}}} = 5a_y^2\\
\end{aligned}\label{eqn2.qo}
\end{equation}
In this form we see that the renormalization group equations of the Yukawa theory have a strong resemblance to the system of differential equations used to describe the $SIR$ model. We may then expect that techniques and visualizations used in the renormalization group equations should be applicable to the equations used in epidemic modeling. Below we illustrate this using the example of flow diagrams and mixing terms in the equations.
\begin{figure}
\centering
  \includegraphics[width =  .5\linewidth]{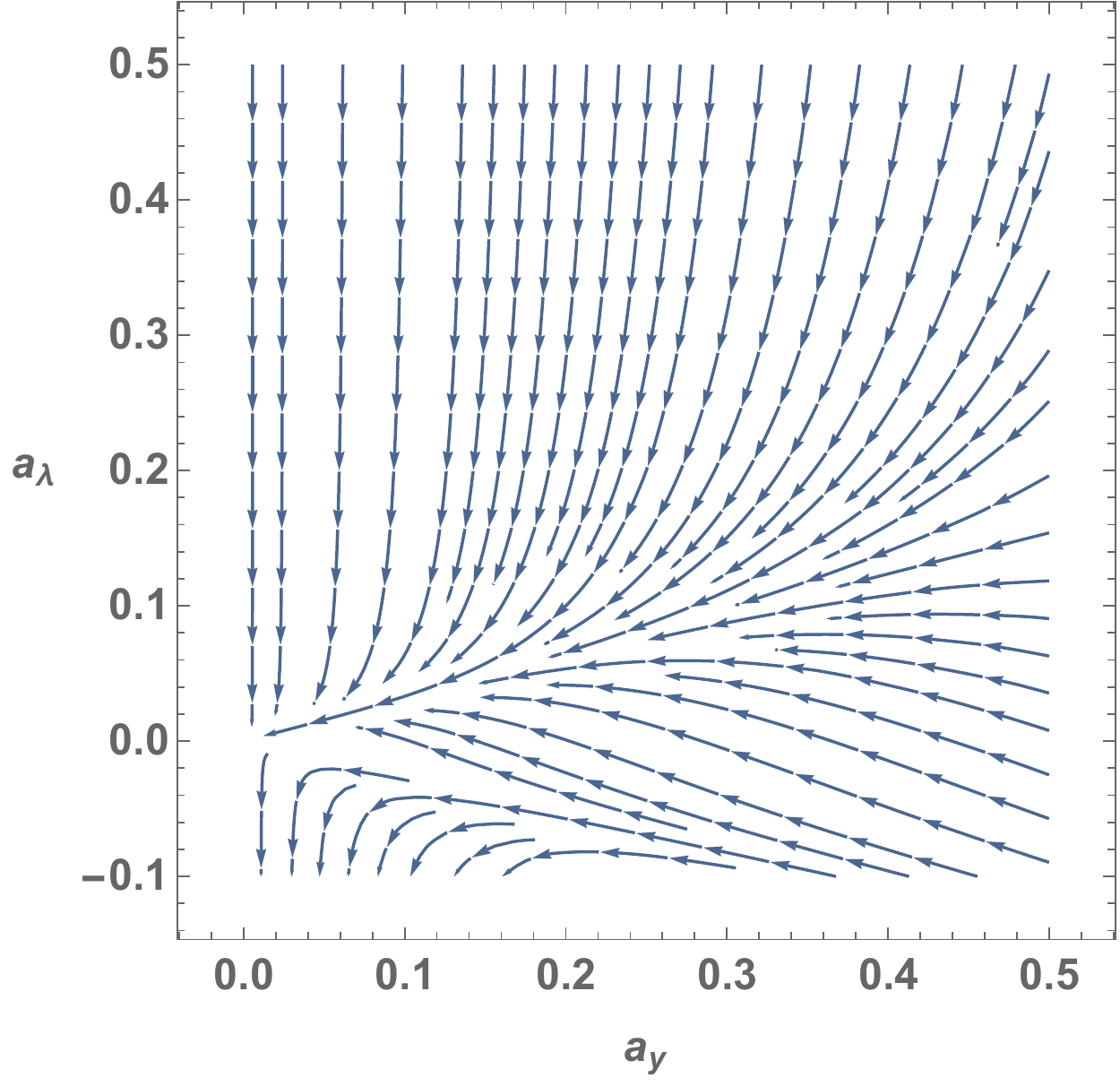}
  \caption{Flow diagram for renoralization group for the Yukawa theory of high energy physics}
  \label{fig:Radion Potential}
\end{figure}

\begin{figure}
\centering
  \includegraphics[width =  .5\linewidth]{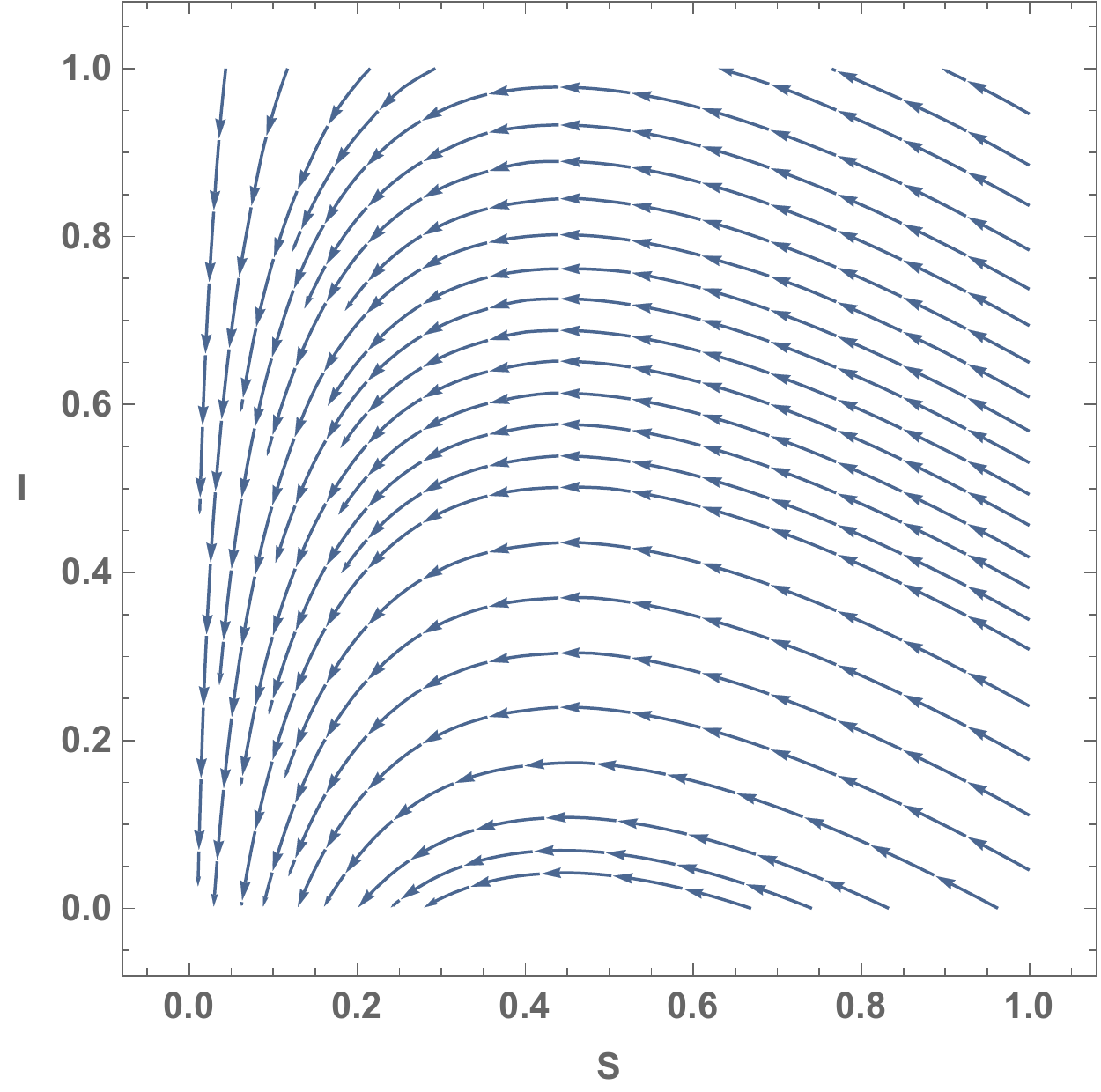}
  \caption{Flow diagram for the $SIR$ model for epidemic modeling. }
  \label{fig:Radion Potential}
\end{figure}

\newpage

\subsection{Flow diagrams for renormalization group and epidemic modeling}

The renormalization group flow diagrams illustrate fixed points in the renormalization flow where the coupling constants do not change in scale. Thus they become fixed at whatever value make the beta functions vanish. In the pandemic context a similar phenomena happens when we reach a point in the pandemic when the number of susceptible, infected and recovered individuals are not changing in time. This can occur if the number of susceptible  individuals dropped to a level that the pandemic is extinguished or it can also occur if the number of infected individuals reaches a constant level in the population and then the pandemic virus becomes endemic in the population. In such cases the fixed points of the flow diagram represent points of endemic equilibrium in $(S,I,R)$ space \cite{Narsingani}\cite{Lucas}. In figure 2 we show the renormalization group flow for the Yukawa theory considered in \cite{Molgaard:2014mqa}. It is clear from the diagram that the flow of coupling constant when the scale changes goes to a fixed point in coupling constant space where the flow is stationary and the beta functions or the right hand side of the renormalization group equations vanishes. In figure 3 we show a similar flow diagram for the $SIR$ epidemic model. In this case we also see that the flow goes to a point where the compartments $I$ and $S$ become stationary when the epidemic has ended after the number of susceptible individuals drops to a level so that the epidemic can no longer be sustained. It will also be important to study flow visualizations over longer time scales where the virus can be come endemic and then recur at a later date.

\subsection{Effect of epedemic contact matrix and  mixing in the renoramalization group equations}


The contact matrix is a symmetric matrix that indicates the number of contacts between different groups. If the matrix is diagonal then the different groups do not mix. For example if two countries are island countries and have no boat or air travel between them then contact matrix will be diagonal. For the renormalization group one has an analogous concept of operator mixing. This occurs because loop graphs induce new operators that were not present in in initial renormalization group evolution. The matrix of coefficients that describe this operator mixing is the anomalous dimension matrix and is analogous to the contact matrix.

For the renormalization group equation  the effect of operator mixing is described by the anomalous dimension matrix $\gamma_{i,j}$ through the equation:
\begin{equation}\frac{{d{C_i}}}{{dt}} = {\gamma _{ij}}(C){C_j}\end{equation}
where $C_i$ is the coefficient of an operator 
$O_i$. These matrices have an important physical effect as operators which have vanishing coefficient at one scale but which develop a nonzero value at a different scale through the mixing with other operators in the theory. 

For the $SIR$ model the contact matrix plays a similar role. In this case a subset population may have a vanishing number of infections at one time but develop infections through the contact with another segment of the population at a later time. This contact matrix is proportional to the number of contacts between the different populations and is represented in the differential equations of the $SIRD$ model in a similar manner to the anomalous dimension matrix in the Renormalization Group Equations. For the $SIRD$ model the effect contact matrix $ c_{ij}$ is given by \cite{Mossong}:
\begin{equation}\begin{aligned}
&\frac{{d{S_i}}}{{dt}} =  - {\beta _i}{S_i}{I_i} - {c_{ij}}{S_i}{I_j}\\
&\frac{{d{I_i}}}{{dt}} = {\beta _i}{S_i}{I_i} - {c_{ij}}{S_i}{I_j} - {\gamma _i}{I_i} - {\mu _i}{I_i}\\
&\frac{{d{R_i}}}{{dt}} = {\gamma _i}{I_i}\\
&\frac{{d{D_i}}}{{dt}} = {\mu _i}{I_i}\\
\end{aligned}\label{eqn2.qo}
\end{equation}
In these equations $\beta_i$ is the contact coefficient between susceptible and infected individuals in group $i$, $\gamma_i$ is the infection rate in group $i$ and $\mu_i$ is the death rate in group $i$ and $c_{ij}$ is the contact matrix between different groups. In the $SIRD$ model $S_i$ is the number of susceptible individuals in group $i$, $I_i$ is the number of infected individulas in group $i$, $R_i$ is the number of recovered individuals in group $i$ and $D_i$ is the number of individuals who died in group $i$. The $SIRD$ model has been applied to pandemics including the ongoing COVID-19 pandemic in \cite{Villaverde}\cite{Chatterjee}\cite{Chakraborty}. We will give two examples of the use of the $SIRD$ model in grouping by age and also by country in the next two sections.





\section{Effect of contact matrix between different age groups in pandemic modeling}

The contact matrix between  different  age groups can be important especially if one age group is more susceptible  or has a higher mortality rate than the other. For example the older population may be more inclined to severe outcomes if they have weakened immune systems or secondary health conditions. The other way around is possible too. Older segments of the population may have been exposed earlier in life to a similar pathogen and still retain some immunity while the younger population would not been alive during this time and would be completely susceptible. Still another scenario is that the younger population was vaccinated against a similar pathogen as part of routine child immunizations wheres the older population either never received them or the effect of the vaccines have waned leaving them vulnerable. The $SIRD$ model among two age groups is given by an eight compartment model \cite{Yin}:
\begin{equation}\begin{aligned}
&\frac{{d{S_1}}}{{dt}} =  - {\beta _1}{S_1}{I_1} - {c_{12}}{S_1}{I_2}\\
&\frac{{d{I_1}}}{{dt}} = {\beta _1}{S_1}{I_1} - {c_{12}}{S_1}{I_2} - {\gamma _1}{I_1} - {\mu _1}{I_1}\\
&\frac{{d{R_1}}}{{dt}} = {\gamma _1}{I_1}\\
&\frac{{d{D_1}}}{{dt}} = {\mu _1}{I_1}\\
&\frac{{d{S_2}}}{{dt}} =  - {\beta _2}{S_2}{I_2} - {c_{21}}{S_2}{I_1}\\
&\frac{{d{I_2}}}{{dt}} = {\beta _2}{S_2}{I_2} - {c_{21}}{S_2}{I_1} - {\gamma _2}{I_2} - {\mu _2}{I_2}\\
&\frac{{d{R_2}}}{{dt}} = {\gamma _2}{I_2}\\
&\frac{{d{D_2}}}{{dt}} = {\mu _2}{I_2}\\
\end{aligned}\label{eqn2.qo}
\end{equation}
The contact matrix between the two age groups is taken to be:
\begin{equation} 
 c_{ij} = \begin{bmatrix}
 
   0 & .075  \\ 
   .075 & 0   \\ 
\end{bmatrix}
  \end{equation}
Other parameters we use are: 
\begin{equation}\begin{aligned}
&(\beta_1=\beta_2 = \frac{3}{2})\\ 
&\gamma_1= .01 (99.834) \frac{2}{3}\\
&\gamma_2= .01 (96.25) \frac{2}{3}\\
&\mu_1= .01 (.166) \frac{2}{3}\\
&\mu_2= .01 (3.75) \frac{2}{3}\\
\end{aligned}\label{eqn2.qo}
\end{equation}
Where group 1 is the age group 0-49 with a low mortality rate and group 2 is the age group 50+ with a substantially higher mortality rate. The initial population of group 1 is taken to be 60,000 and for group 2 the initial population taken to be 40,000 so that the total initial population of the two groups is 100,000. The initial number of infected individuals in group 1 was taken to be 100 and in group 2 it was taken to be 0. Solving these equations with the parameters and initial condition we see that as the infections in group 1 go down but once contact with group 2 is initiated the infections in group 2 increase and one obtains a second peak of infections in group 2 with contributes substantially  to the total number of deaths due to the high mortality in group 2.


\begin{figure}[!htb]
\centering
\minipage{0.32\textwidth}
  \includegraphics[width=\linewidth]{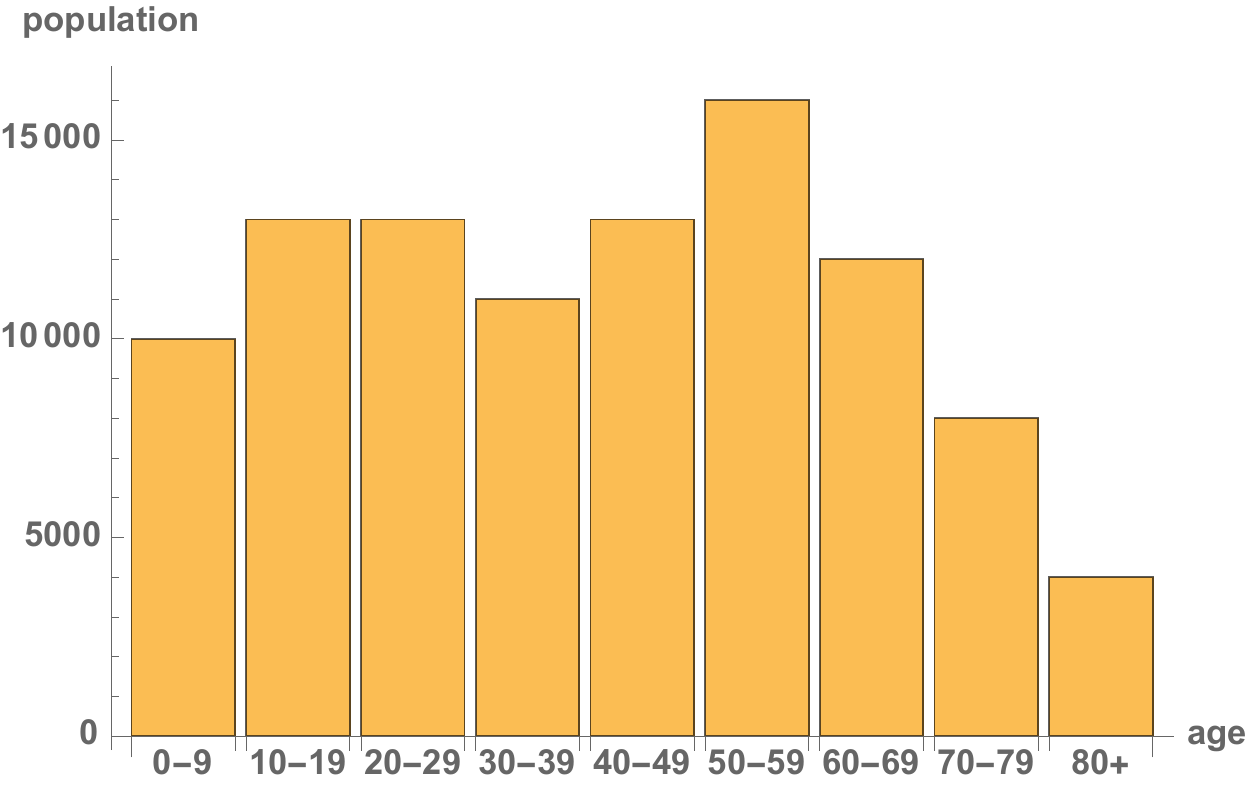}
\endminipage\hfill
\minipage{0.32\textwidth}
  \includegraphics[width=\linewidth]{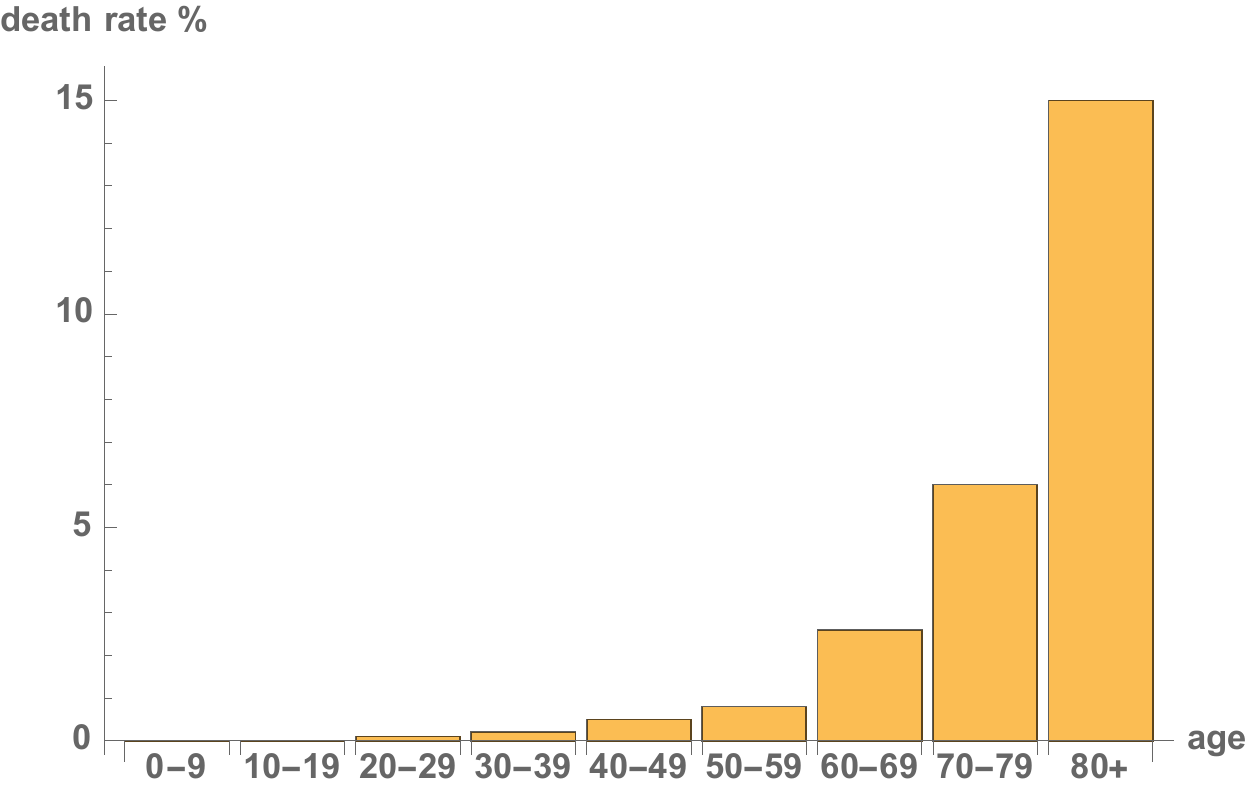}
\endminipage\hfill
\minipage{0.32\textwidth}%
  \includegraphics[width=\linewidth]{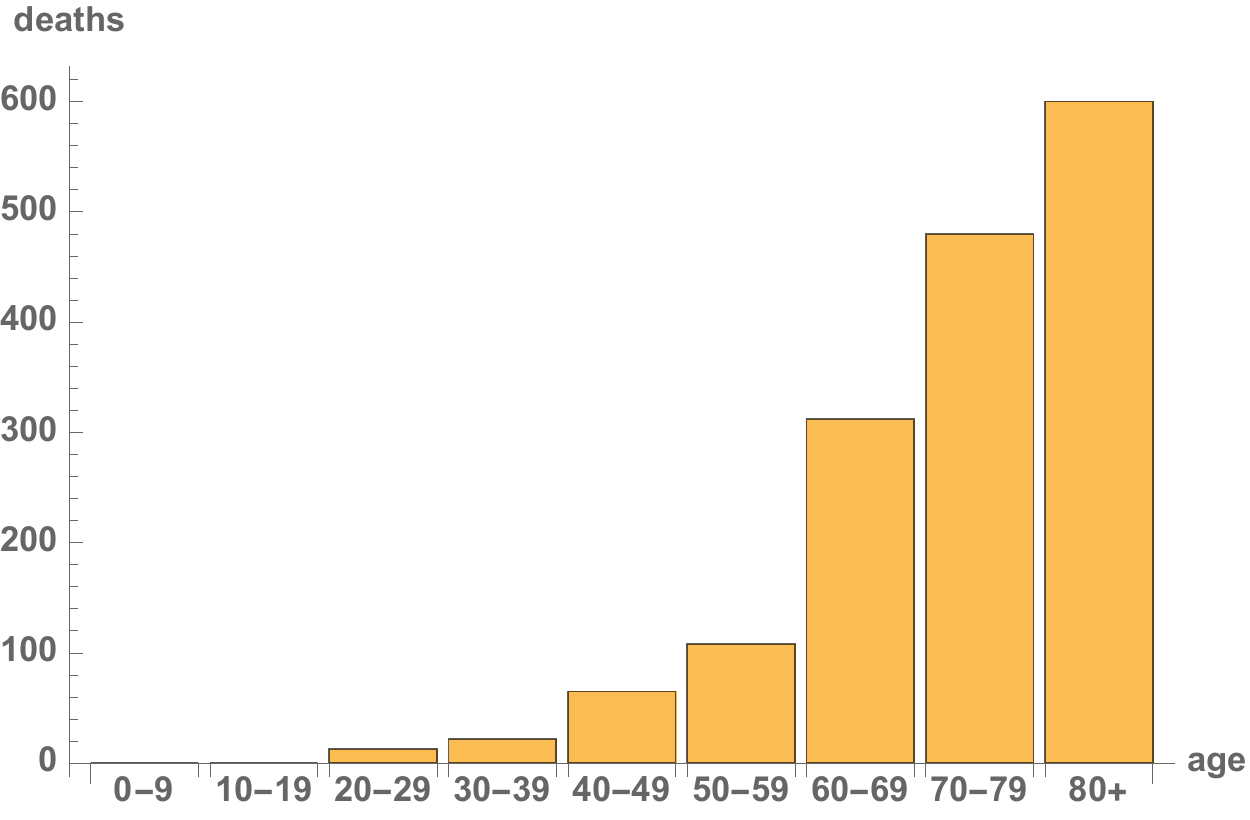}
 
\endminipage
\caption{(Left) Typical age distribution of a population of of 100,000.(Middle) Death rate as a function of age for a pandemic with characteristics similar to the COVID-19 pandemic (Right) Expected number of deaths in an  infected population of population 100,000 for a pandemic of similar characteristics to COVID-19 pandemic.}
\end{figure}

\begin{figure}[!htb]
\minipage{0.32\textwidth}
  \includegraphics[width=\linewidth]{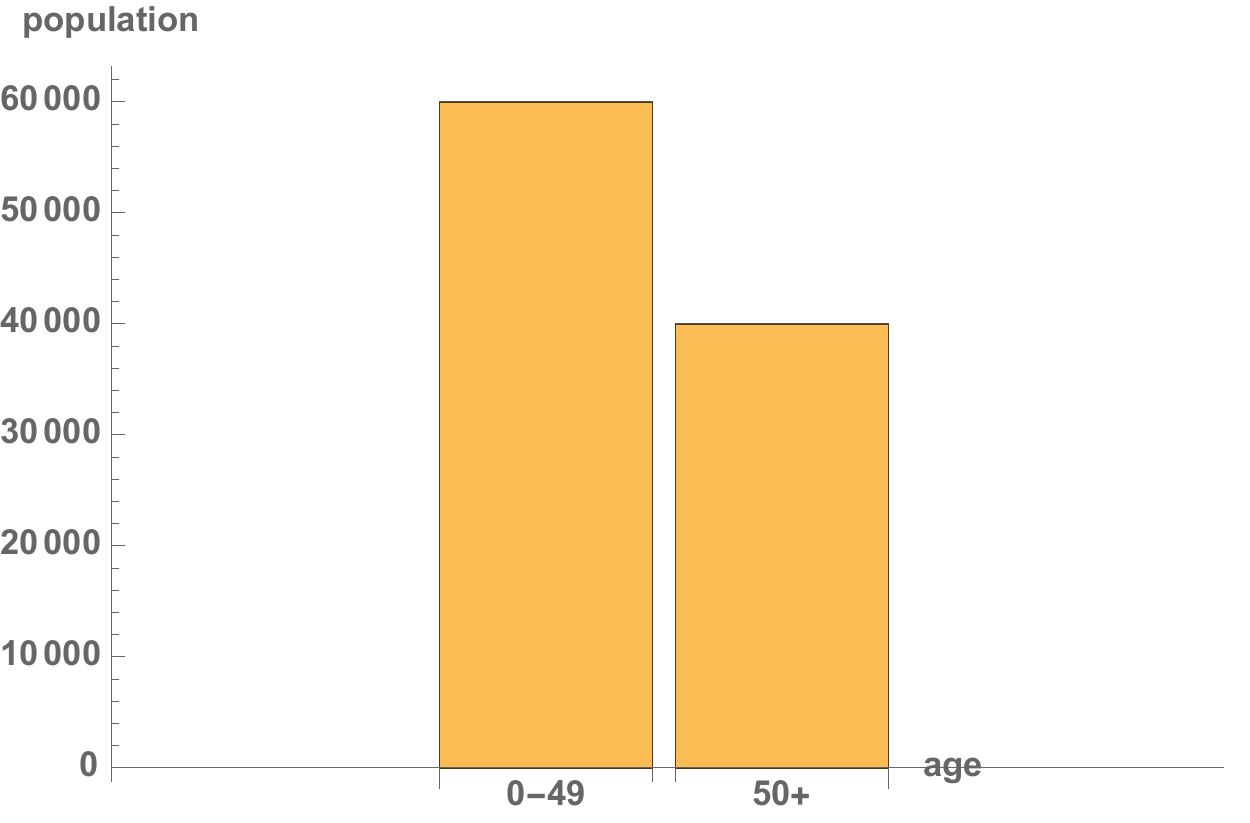}
\endminipage\hfill
\minipage{0.32\textwidth}
  \includegraphics[width=\linewidth]{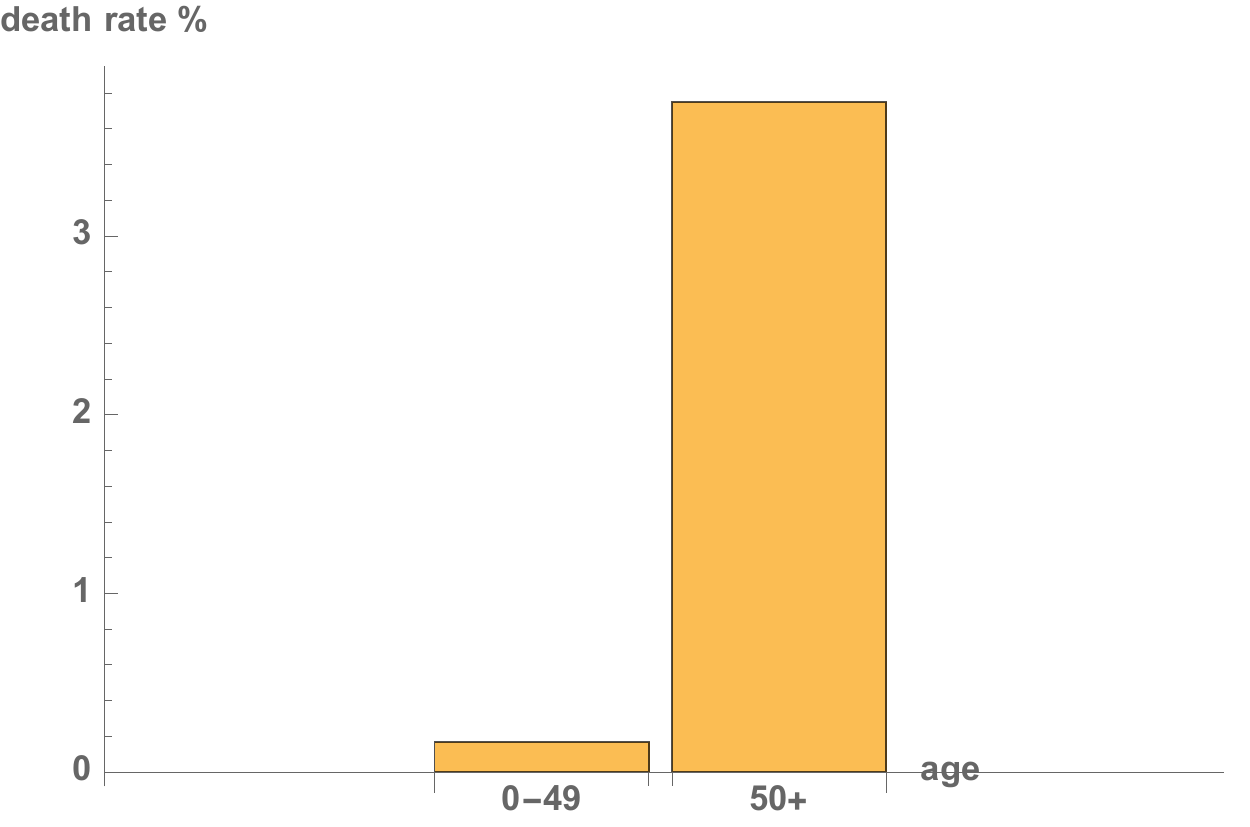}
\endminipage\hfill
\minipage{0.32\textwidth}%
  \includegraphics[width=\linewidth]{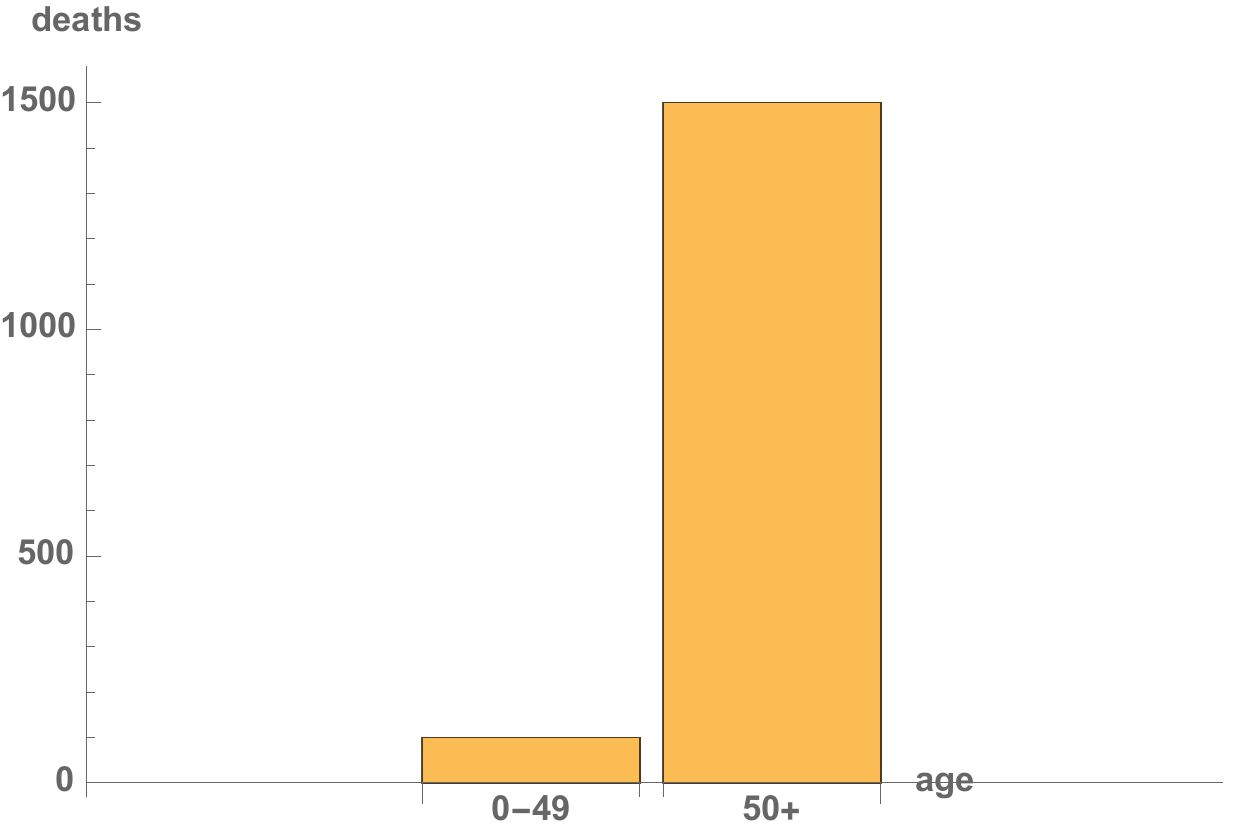}
 
\endminipage
\caption{(Left) Typical age distribution for two age groups $0-49$ and $50+$ of a population of of 100,000.(Middle) Death rate as a function of age for a pandemic with characteristics similar to the COVID-19 pandemic (Right) Expected number of deaths in an  infected population of population 100,000 for a pandemic of similar characteristics to COVID-19 pandemic.}
\end{figure}


\begin{figure}
\centering
  \includegraphics[width =  \linewidth]{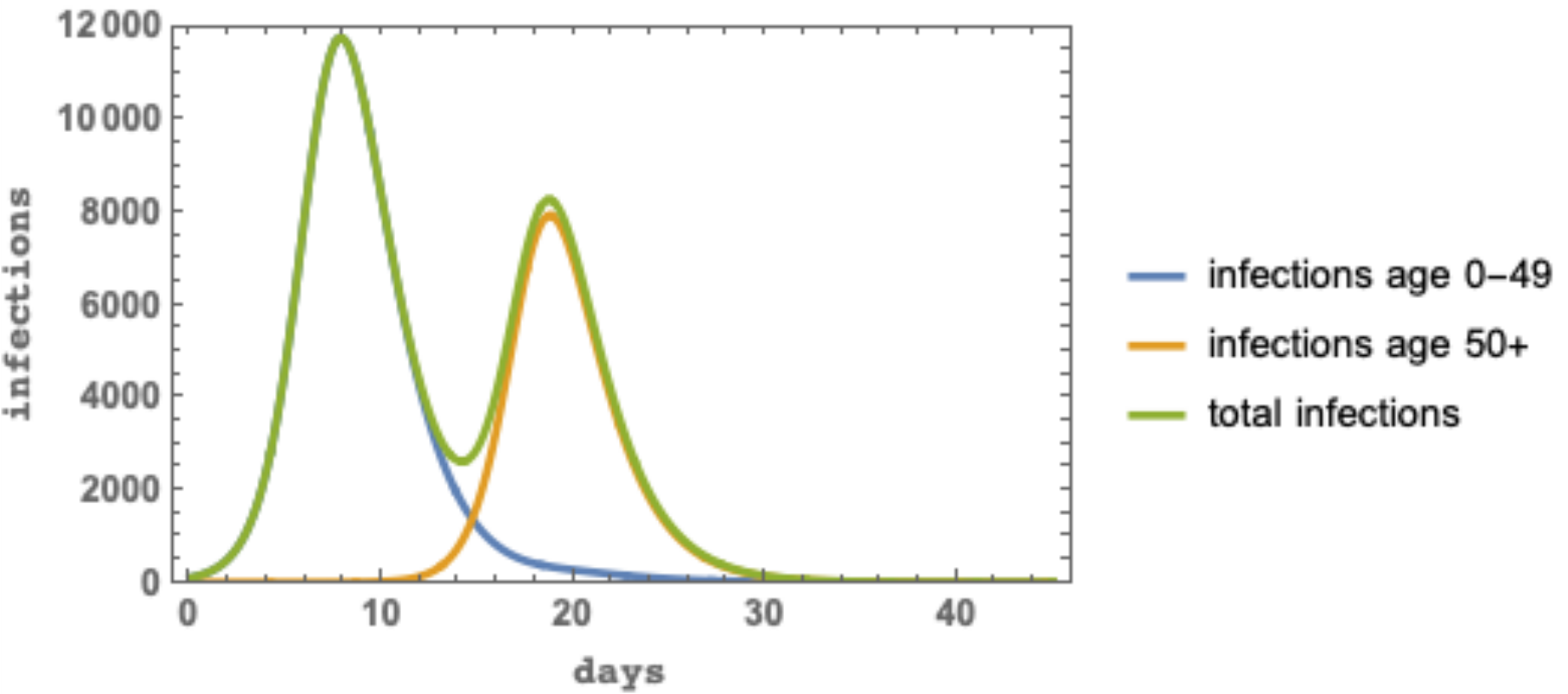}
  \caption{Infections as a function of time for two age groups $(0-49)$ and $50+$ where there is a nondiagonal contact matrix in the SIRD model describing the interaction between the two age groups during the pandemic. Note the presence of two peaks corresponding to peak infections in both age groups. }
  \label{fig:Radion Potential}
\end{figure}

\begin{figure}
\centering
  \includegraphics[width =  \linewidth]{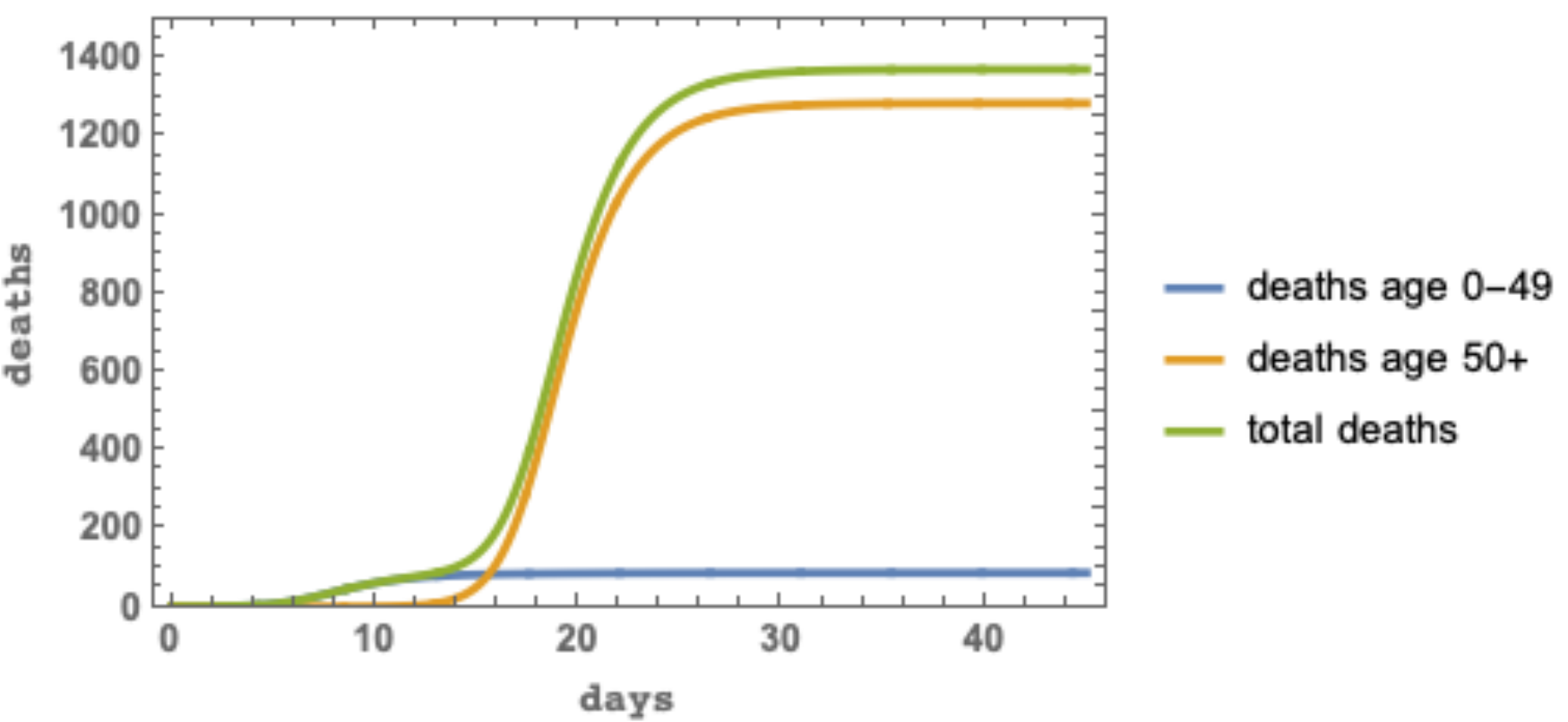}
  \caption{Deaths as a function of time for two age groups $(0-49)$ and $50+$ where there is a nondiagonal contact matrix in the SIRD model describing the interaction between the two age groups during the pandemic. Note the older age group leads to far more deaths than the younger age group due to the higher mortality rate in the older age group.}
  \label{fig:Radion Potential}
\end{figure}

Often during pandemics segments of the population are kept from the susceptible part of the population through stay at home orders. These segments  may be added later to susceptible pool at a later stage in the pandemic when restrictions are eased. This can be modeled with nine groups from $0$ to $80+$ in ten year intervals as in figure 4 or as two age groups one group under 50 and the other over 50 as in figure 5. In either case the population with a contact matrix which is diagonal initially and is non-diagonal at a later stage. The results from the $SIRD$ model for the the two age group case is shown in figure 6 and figure 7. We clearly see the second peak of infection in the older age group due to the the effects of the contact matrix and initial infection in the younger age group. Because of the higher mortality rate in the older age group the model predicts that most of the deaths occur there. For the parameters that we chose in the $SIRD$ simulation with a susceptible population of 60,000 in the younger age group and 40,000 in the older age group and we found that the number of deaths were 86 in the younger age group and 1283 in the older age group for at total of 1369 deaths out of the 100,000 population. This  is a dramatic illustration of the effect of the contact matrix between different age populations in pandemic modeling. 

It is important to note in the early phase of a pandemic such as from COVID-19 the model parameters are not known accurately. These uncertainties can lead to  a range of predictions that depend on these uncertainties. Although there are several parameters in the models we simulate we found the parameter $\beta$ that indicated the interaction between the susceptible and infected populations to be very important. For the simulation results shown figure 6  increasing $\beta$ by 10 percent lead to a peak in infection in the younger group increasing from 12,000 to 13,800 and and an increase the peak in the older group from 8000 to 9500. Decreasing $\beta$ by ten percent lead to a peak in infection in the younger group with a decrease from 12,000 to 9500 and and an decrease in  the peak in the older group from 8000 to 7060. For the simulation results in figure 7 increasing $\beta$ by 10 percent lead to deaths in the younger group increasing from 86 to 89 and and an increase in deaths in the older group from 1283 to 1335. Decreasing $\beta$ by ten percent lead to deaths in the younger to decrease from 86 to 82 and and an increase in  deaths in the older group from 1283 to 1212.

\newpage

\section{Effect of contact matrix between different countries in pandemic modeling }

In this section we consider a simple three compartment model involving three countries involved in virus spreading. In country 1 the virus starts an exponential growth pattern. Travel between country 1 and country 2 takes place, followed at a later date by travel from country 2 to 3. The three country model can be described by the twelve compartment $SIRD$ model:
\begin{equation}
\begin{aligned}
&\frac{{d{S_1}}}{{dt}} =  - {\beta _1}{S_1}{I_1} - {c_{12}}{S_1}{I_2}\\
&\frac{{d{I_1}}}{{dt}} = {\beta _1}{S_1}{I_1} + {c_{12}}{S_1}{I_2} - {\gamma _1}{I_1} - {\mu _1}{I_1}\\
&\frac{{d{R_1}}}{{dt}} = {\gamma _1}{I_1}\\
&\frac{{d{D_1}}}{{dt}} = {\mu _1}{I_1}\\
&\frac{{d{S_2}}}{{dt}} =  - {\beta _2}{S_2}{I_2} - {c_{21}}{S_2}{I_1}-{c_{23}}{S_2}{I_3}\\
&\frac{{d{I_2}}}{{dt}} = {\beta _2}{S_2}{I_2} + {c_{21}}{S_2}{I_1}  + {c_{23}}{S_2}{I_3}- {\gamma _2}{I_2} - {\mu _2}{I_2}\\
&\frac{{d{R_2}}}{{dt}} = {\gamma _2}{I_2}\\
&\frac{{d{D_2}}}{{dt}} = {\mu _2}{I_2}\\
&\frac{{d{S_3}}}{{dt}} =  - {\beta _3}{S_3}{I_3} - {c_{32}}{S_3}{I_2}\\
&\frac{{d{I_3}}}{{dt}} = {\beta _3}{S_3}{I_3} + {c_{32}}{S_3}{I_2} - {\gamma _3}{I_3} - {\mu _3}{I_3}\\
&\frac{{d{R_3}}}{{dt}} = {\gamma _3}{I_3}\\
&\frac{{d{D_3}}}{{dt}} = {\mu _3}{I_3}\\
\end{aligned}
\label{eqn2.qo}
\end{equation}
with contact matrix between the countries given by the three by three matrix which we take to be:
\begin{equation} 
 c_{ij} = \begin{bmatrix}
 
   0 & .075 & 0 \\ 
   .075 & 0 & .075  \\ 
   0 & .075 & 0   \\ 
\end{bmatrix}
  \end{equation}
Other parameters we use in the simulation are: 
\begin{equation}\begin{aligned}
&(\beta_1=\beta_2=\beta_3= \frac{3}{2})\\ &(\gamma_1=\gamma_2=\gamma_3= .01 (98.4) \frac{2}{3})\\
&(\mu_1=\mu_2=\mu_3= .01 (1.6) \frac{2}{3})\\
\end{aligned}\label{eqn2.qo}
\end{equation}
With these parameters and also multiplying the contact matrix with a smoothed out step function to model time delay in establishing contact between the countries we obtain the results for infections in figure 8 and deaths figure 9. Note that when the first country has contact with the second the infections in country 1 have dropped substantially nevertheless there were enough cases to continue the pandemic into country 2. The same scenario then repeats itself after country 2's infections go down and contact starts between country 2 and country 3. The same step like effect can be seen in deaths in figure 9 where deaths in country 2 pick up after contact is initiated with the initial infection in country 1 with the same scenario continuing between country 2 and 3. Note the three separate peaks in the number of infections which arise when travel restrictions are eased between countries and the stair like structure in the number of deaths as the pandemic moves between the countries.  The number of deaths occurring assuming each country has a population of 100,000 is 1356 per country for a total 4068 deaths for the total population of the three countries of 300,000. This is also a dramatic illustration of the effect of contact matrix in this case between countries in pandemic modeling. 

Again it is important to note in the early phase of a pandemic such from COVID-19 the model parameters are not known accurately. For the simulation results shown figure 8  increasing $\beta$ by 10 percent lead to a peak in infection in the first country increasing from 19,560 to 23,020, an increase the peak in the second country from 19,630 to 23,030, and an increase the peak in the third country from 19,630 to 23,025 . Decreasing $\beta$ by ten percent lead to a peak in infection in the first country with a decrease from 19,960 to 18,800 , a decrease in  the peak in the second country from 19,630 to 16,000 and  a decrease in  the peak in the third country from 19,630 to 16,000 as well. For the simulation results in figure 9 increasing $\beta$ by 10 percent lead to deaths in the first country increasing from 1350 to 1445, an increase in deaths in the second country from 1387 to 1445, and an increase in the deaths in the third country from 1387 to 1417. Decreasing $\beta$ by ten percent lead to deaths in the first country to decrease from 1350 to 1323, deaths in the second country to decrease from 1387 to 1324, and deaths in the third country to decrease from 1387 to 1324 as well. 
\begin{figure}
\centering
  \includegraphics[width =  \linewidth]{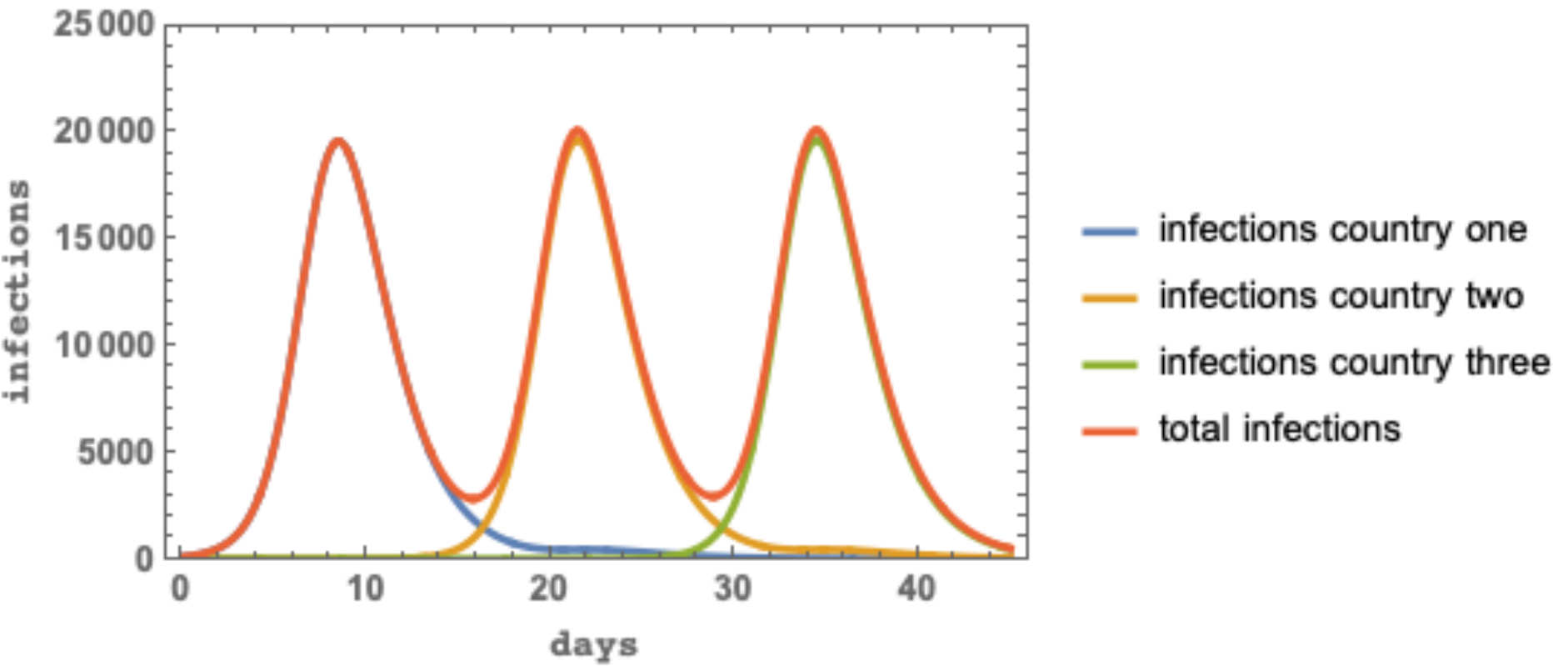}
  \caption{Infections as a function of time for three countries whose interactions are described by an nondiagonal contact matrix. Note that although the contacts between country 1 and country 3 are zero country 3 nevertheless becomes part of the pandemic because both countries 1 and 3 interact with country 2.}
  \label{fig:Radion Potential}
\end{figure}

\begin{figure}
\centering
  \includegraphics[width =  \linewidth]{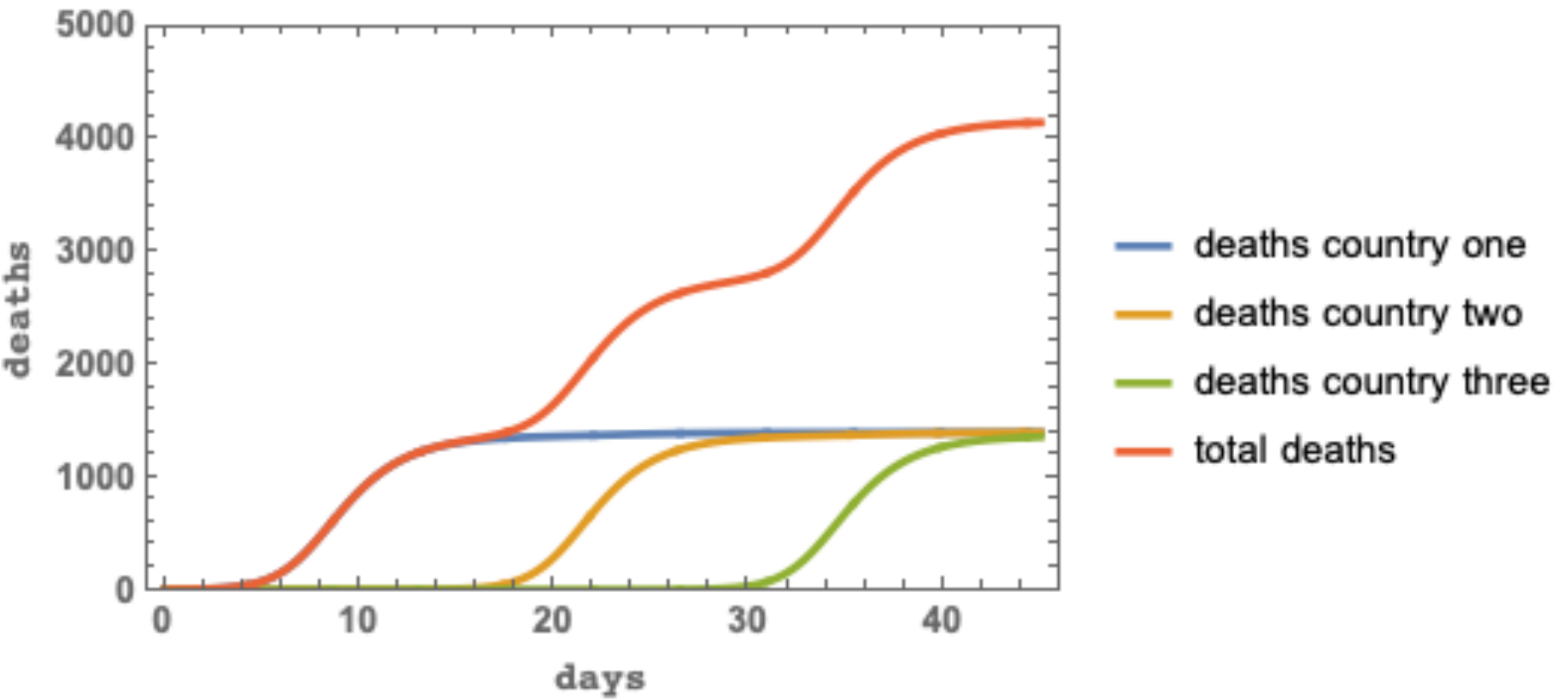}
  \caption{Deaths as a function of time for three countries whose interactions are described by an nondiagonal contact matrix. Note the stair like structure in the number of deaths as the pandemic moves from country 1 to country 2 and from country 2 to country 3.}
  \label{fig:Radion Potential}
\end{figure}

\section{Vaccine waning and indirect effects of nonspecific vaccines}

There are two possible effects of vaccine waning in pandemic modeling. In the first case a vaccine is developed specifically targeting the virus. Like the seasonal flu vaccine the effectiveness of this vaccine can effectively wane perhaps in short enough time scales so that people will lose immunity to the specific virus during the pandemic and will require a booster to regain immunity \cite{Gu}\cite{Hefferman}\cite{Lewnard}\cite{Cohen}\cite{Knipl2}\cite{Ehrhardt}\cite{Feng}\cite{Kasia}. In the second case one sees some efficacy from a vaccine which we call
nonspecific to the pandemic virus but was developed for a different pathogen but nevertheless may have some positive effects that result in lower morbidity to the pandemic virus. In references  \cite{Fidel}\cite{Chumakov}\cite{Escobar} nonspecific vaccines that are being studied including the MMR vaccine, the live attenuated Polio vaccine and the BCG Tuberculosis vaccine. The effects of this nonspecific vaccine  could also wane over time. This can also lead to sporadic outbreaks of the original virus it was designed to protect against as for example has been seen in outbreaks of mumps. If the nonspecific vaccine has a positive effect against the pandemic then the effect of vaccine waning may also contribute to age dependence of pandemic morbidity as vaccine waning may reduce the effect of the nonspecific vaccine  in older populations who have not received a booster and have a longer time period from inoculation. The waning involved in the nonspecific vaccine  could be much longer than the time frame of the pandemic and the main effect would be a contribution to the age distribution of pandemic morbidity.

Here we study a  set of two $SIRD$ models to study this effect. In the first model we study a $SIRD$ model with two age groups and no effect of the nonspecific vaccine. 
\begin{equation}\begin{aligned}
&\frac{{d{S_1}}}{{dt}} =  - {\beta _1}{S_1}{I_1} - {c_{12}}{S_1}{I_2}\\
&\frac{{d{I_1}}}{{dt}} = {\beta _1}{S_1}{I_1} - {c_{12}}{S_1}{I_2} - {\gamma _1}{I_1} - {\mu _1}{I_1}\\
&\frac{{d{R_1}}}{{dt}} = {\gamma _1}{I_1}\\
&\frac{{d{D_1}}}{{dt}} = {\mu _1}{I_1}\\
&\frac{{d{S_2}}}{{dt}} =  - {\beta _2}{S_2}{I_2} - {c_{21}}{S_2}{I_1}\\
&\frac{{d{I_2}}}{{dt}} = {\beta _2}{S_2}{I_2} - {c_{21}}{S_2}{I_1} - {\gamma _2}{I_2} - {\mu _2}{I_2}\\
&\frac{{d{R_2}}}{{dt}} = {\gamma _2}{I_2}\\
&\frac{{d{D_2}}}{{dt}} = {\mu _2}{I_2}\\
\end{aligned}\label{eqn2.qo}
\end{equation}
In the second model we consider a $SIRD$ model including a possible effect reducing the morbidity with a greater effect on the younger age group due to possible waning of the nonspecific vaccine. The amount of waning by age group is given in figure 10 and the effect on the death rate with or without the hypothetical nonspecific vaccine that we used in the simulation are shown in figure 11. The model we used for the second $SIRD$ model is given by:
\begin{equation}\begin{aligned}
&\frac{{d{S_1}}}{{dt}} =  - {\beta _1}{S_1}{I_1} - {c_{12}}{S_1}{I_2}\\
&\frac{{d{I_1}}}{{dt}} = {\beta _1}{S_1}{I_1} - {c_{12}}{S_1}{I_2} - {\gamma _1}{I_1} - \mu_{v1}{I_1}\\
&\frac{{d{R_1}}}{{dt}} = {\gamma _1}{I_1}\\
&\frac{{d{D_1}}}{{dt}} = \mu _{v1}{I_1}\\
&\frac{{d{S_2}}}{{dt}} =  - {\beta _2}{S_2}{I_2} - {c_{21}}{S_2}{I_1}\\
&\frac{{d{I_2}}}{{dt}} = {\beta _2}{S_2}{I_2} - {c_{21}}{S_2}{I_1} - {\gamma _2}{I_2} - \mu _{v2}{I_2}\\
&\frac{{d{R_2}}}{{dt}} = {\gamma _2}{I_2}\\
&\frac{{d{D_2}}}{{dt}} = \mu _{v2}{I_2}\\
\end{aligned}\label{eqn2.qo}
\end{equation}
Then we compare the results of these two models to see the effect of the nonspecific vaccine and the pandemic. We see from figure 12 that the hypothetical nonspecific vaccine had a positive effect on the mortality of the pandemic with a larger effect on the younger age group. Parameters we used in this simulations are:
\begin{equation}\begin{aligned}
&(\beta_1=\beta_2 = \frac{3}{2})\\ 
&\gamma_1= .01 (99.834) \frac{2}{3}\\
&\gamma_2= .01 (96.25) \frac{2}{3}\\
&\mu_1= .01 (.305) \frac{2}{3}\\
&\mu_2= .01 (3.95) \frac{2}{3}\\
&\mu_{v1}= .01 (.166) \frac{2}{3}\\
&\mu_{v2}= .01 (3.75) \frac{2}{3}\\
\end{aligned}\label{eqn2.qo}
\end{equation}
with the same contact matrix $c_{ij}$ from (4.2).
We see that there is an age related effect in pandemic modeling somewhat similar to what is seen in the COVID-19 pandemic. It would be very interesting add more compartments and to represent comorbidities in the different age groups to more closely model the different age populations reaction to the disease.

\begin{figure}[!htb]
\centering
\minipage{0.5\textwidth}
  \includegraphics[width=\linewidth]{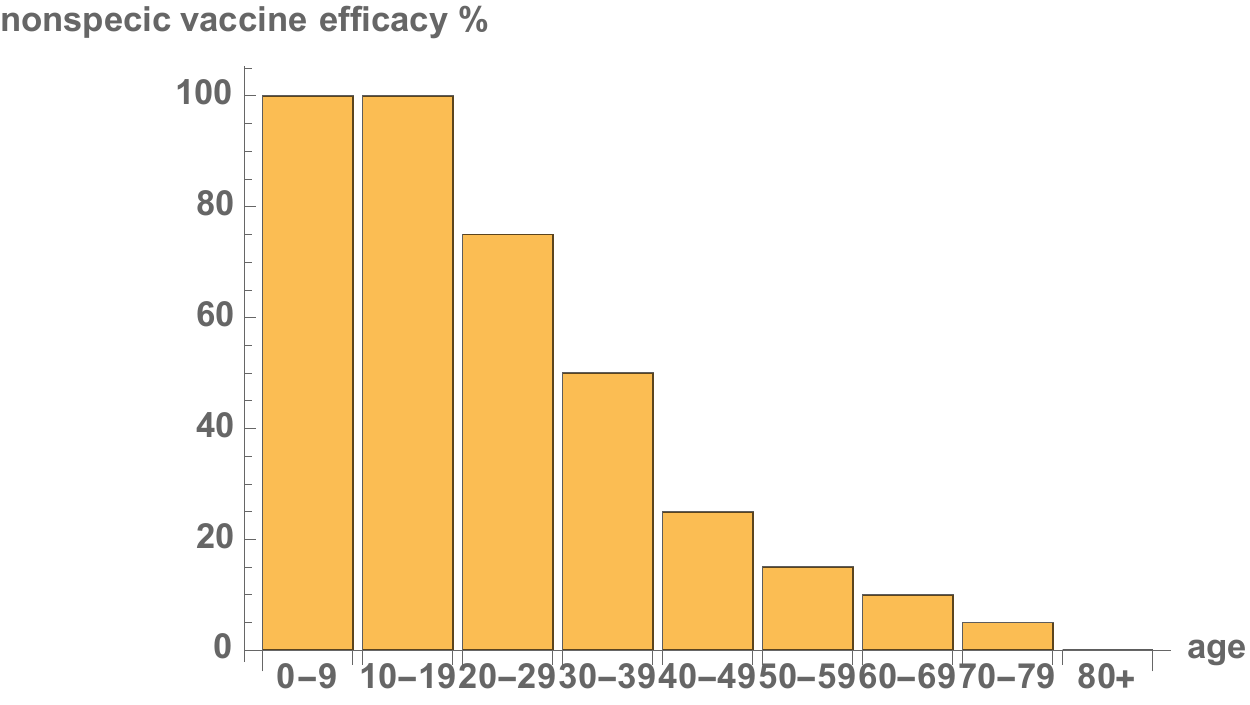}
\endminipage\hfill
\minipage{0.5\textwidth}
  \includegraphics[width=\linewidth]{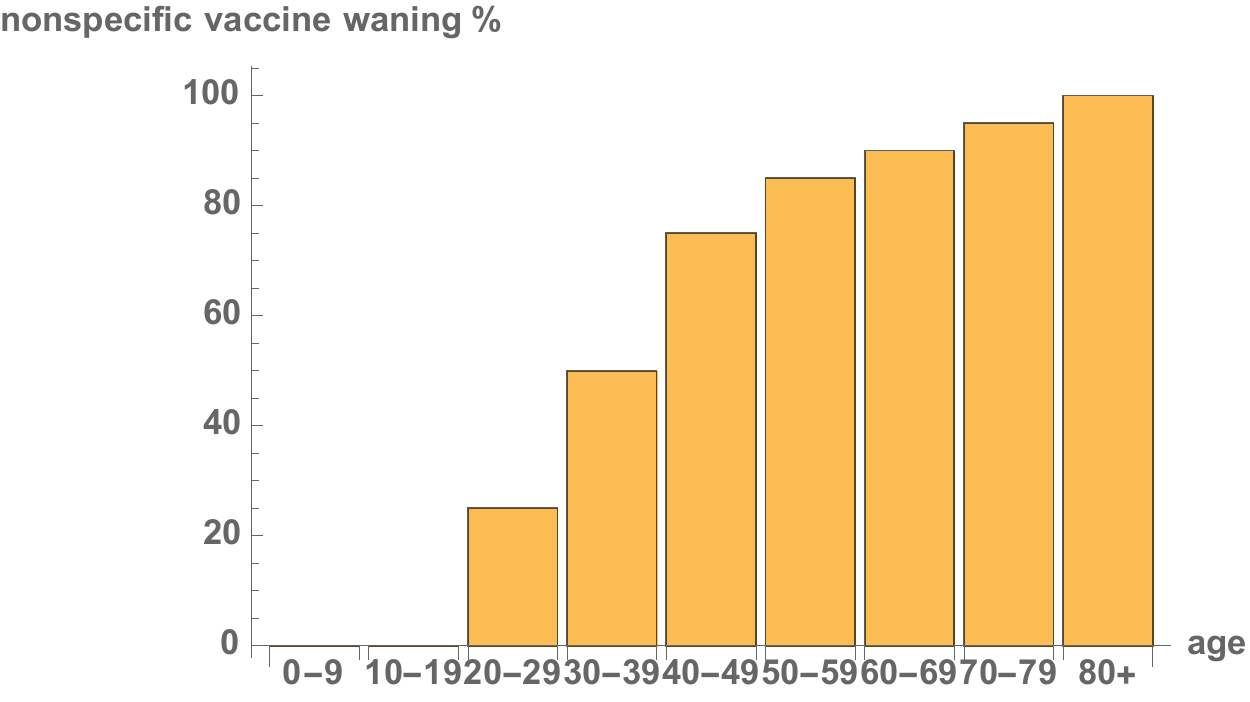}
\endminipage\hfill
\caption{(Left) Vaccine efficacy for a hypothetical nonspecific vaccine which reduces mortality for the pandemic pathogen. Note the efficacy goes to zero in the $80+$ age group if the effect nonspecific  vaccine completely wanes away. (Right) Vaccine waning for a hypothetical nonspecific vaccine which reduces mortality for the pandemic pathogen. Note the waning goes to $100 \%$ in the $80+$ age group if the effect nonspecific  vaccine completely wanes away.}
\end{figure}

\begin{figure}[!htb]
\centering
\minipage{0.5\textwidth}
  \includegraphics[width=\linewidth]{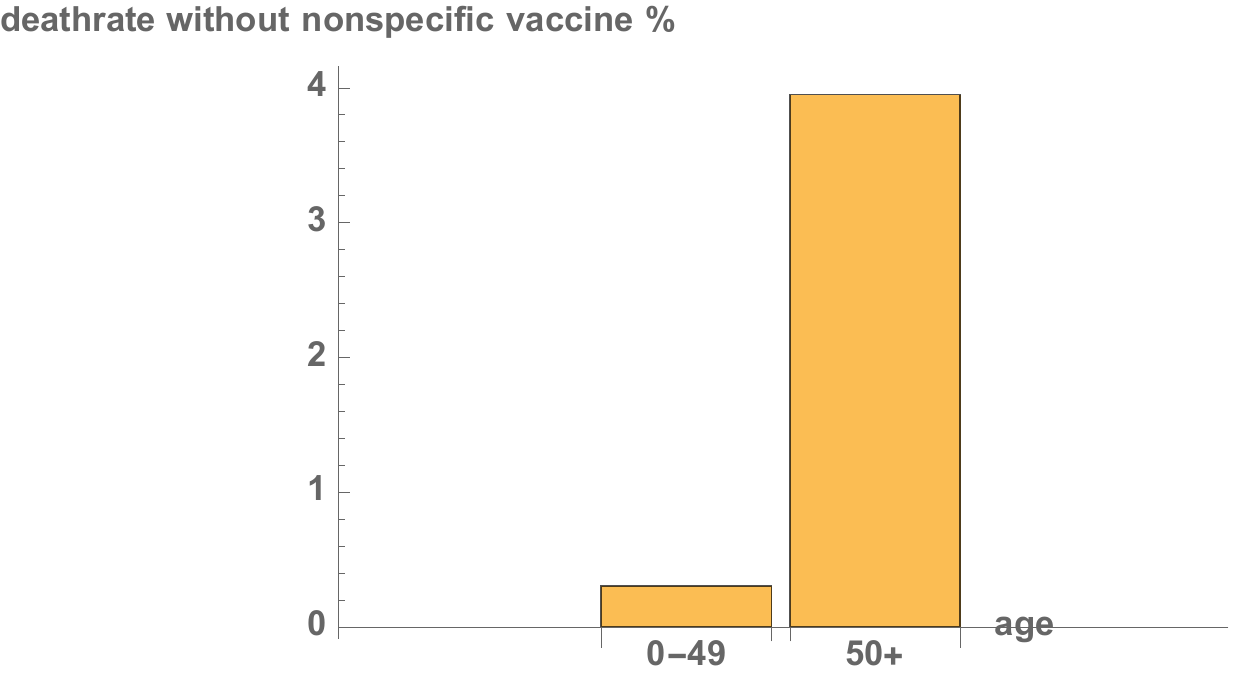}
\endminipage\hfill
\minipage{0.5\textwidth}
  \includegraphics[width=\linewidth]{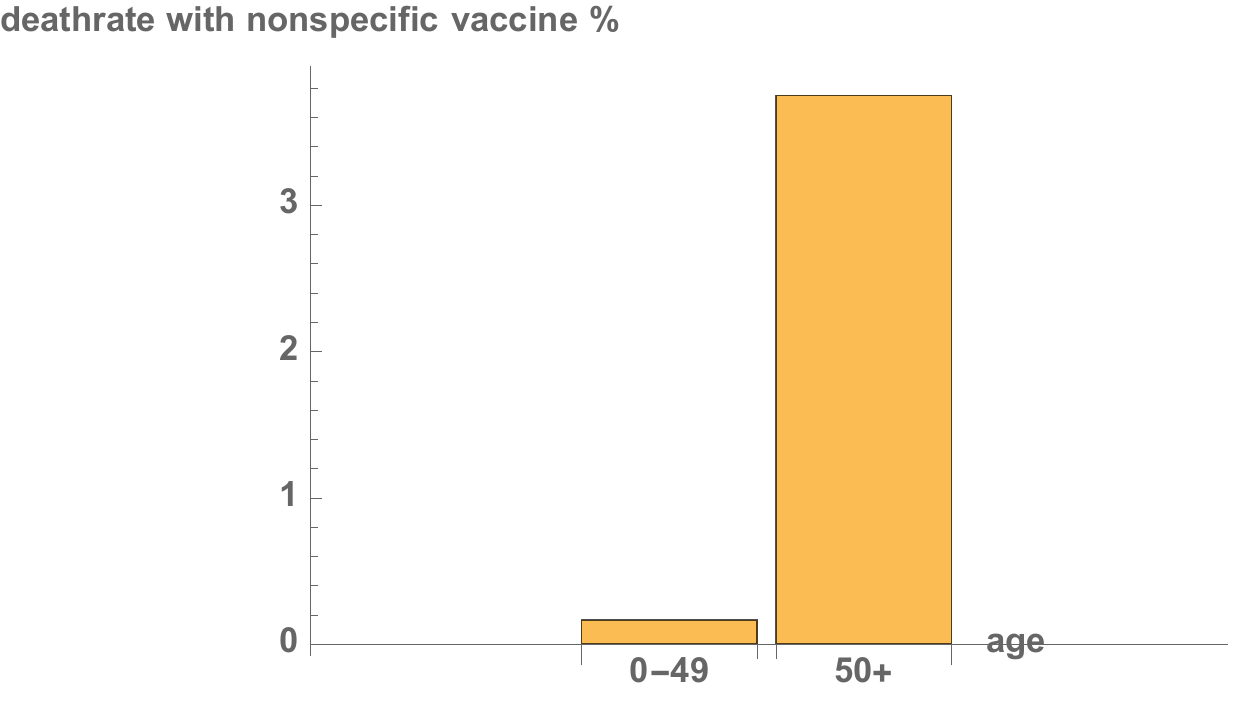}
\endminipage\hfill
\caption{(Left) Death rate for a population without a nonspecific vaccine for two age groups $(0-49)$ and $50+$. (Right) Deathrate for a population with a nonspecific vaccine taking into account vaccine waning for two age groups $(0-49)$ and $50+$. Note the deathrate is reduced more for the younger age group as the nonspecific vaccine wanes less than the older age group.}
\end{figure}

\begin{figure}[!htb]
\centering
\minipage{0.5\textwidth}
  \includegraphics[width=\linewidth]{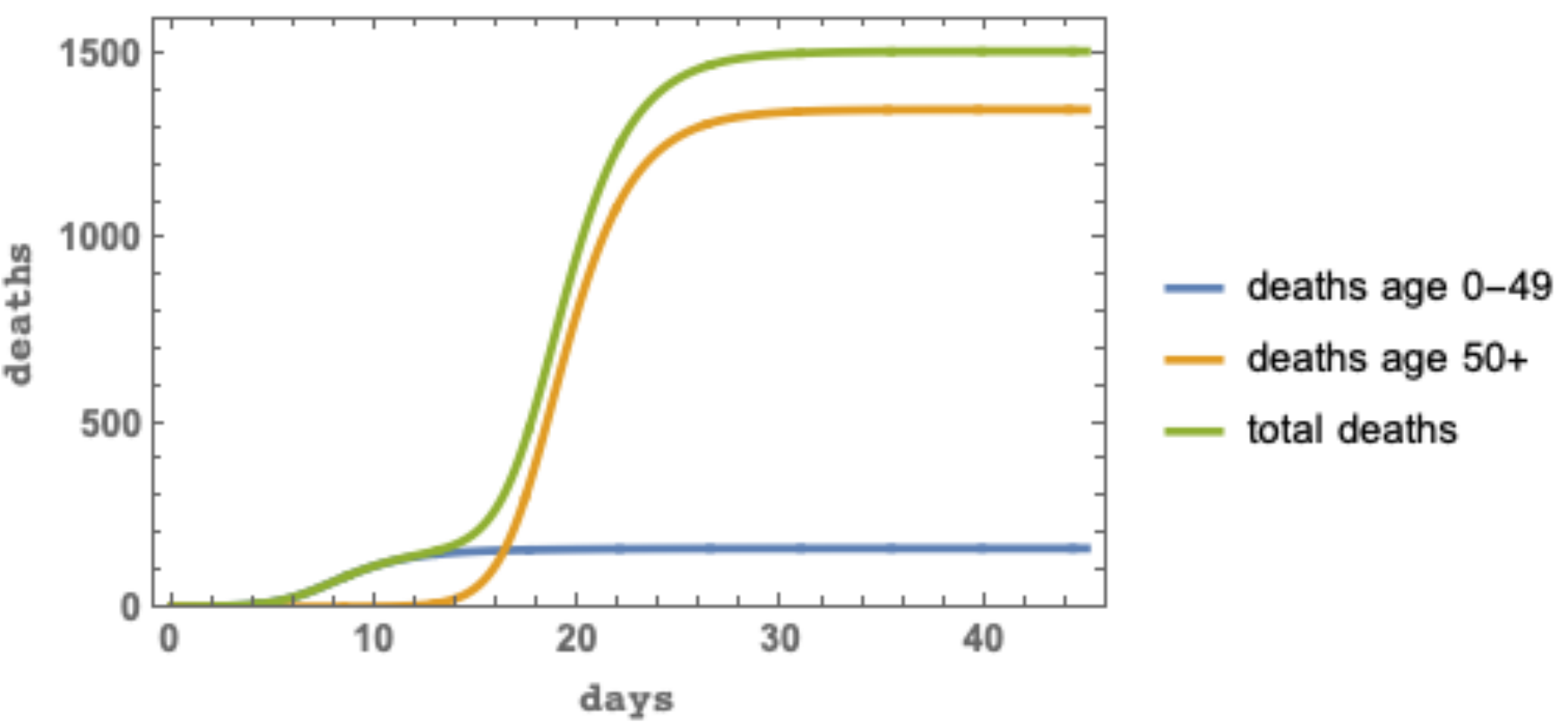}
\endminipage\hfill
\minipage{0.5\textwidth}
  \includegraphics[width=\linewidth]{deatheagegroup.pdf}
\endminipage\hfill
\caption{(Left) Deaths  for a population without a hypothetical nonspecific vaccine for two age groups $(0-49)$ and $50+$. (Right) Deaths for a population with a hypothetical nonspecific vaccine taking into account vaccine waning for two age groups $(0-49)$ and $50+$. Note the number of deaths is reduced more for the younger age group as the nonspecific vaccine wanes less than in the older age group.}
\end{figure}

We also see from figure 12 that the number of deaths in both age groups decrease with the hypothetical nonspecific vaccine. We chose parameters in the $SIRD$ simulation from (6.3) with a susceptible population of 60,000 in the younger age group and 40,000 in the older age group. We found that the number of deaths decreased from 158 to 86 with the use of the hypothetical nonspecific vaccine in the younger age group. The number of deaths in the older age group decreased from 1351 to 1283 with the use of the hypothetical nonspecific vaccine. Overall the number of deaths in the total population of 100,000 decreased from 1509 to 1369 due to the effect of the hypothetical nonspecific vaccine using the parameters (6.3) of the $SIRD$ model. The effect in the older age group is less because of the vaccine waning varying with the number of years from vaccination. The application of a booster of the nonspecific vaccine to the older age group could be expected to lower the mortality even further as more fatalities are recorded in the older age group where the mortality rate is higher. It will be interesting to examine the projections for pandemic modeling as more data becomes available on the effect of booster for nonspecific vaccines \cite{Fidel}\cite{Chumakov}\cite{Escobar}. This type of analysis could also be important for specific vaccines as more data becomes available from clinical trials and measurements are made as to the amount waning that occurs for vaccines designed specifically for the pandemic pathogen \cite{Folegatti}.

Finally The number of infections per day remain the same with or without the nonspecific vaccines in the $SIRD$ model we considered. This is because the nonspecific vaccines reduce the severity of the illness and lower mortality but do not prevent infection with the pandemic pathogen as would happen with a specific vaccine targeted to the pandemic virus \cite{Folegatti}.

For early stages of pandemics model parameters are not known accurately. To gauge the effects of parameter uncertainties  in figure 12 we vary the important parameter $\beta$ to see its effects on the simulation. For figure 12 (left) increasing $\beta$ by 10 percent increased the deaths in the simulation without the nonspecific vaccine from 158 to 164 for the younger age group and from 1351 to 1406 for the older age group. Decreasing $\beta$ by 10 percent decreased the deaths in the simulation without the nonspecific vaccine in the younger group from 158 to 150 and from 1351 to 1277 in the older age group. For figure 12 (right) increasing $\beta$ by 10 percent increased the deaths in the simulation with the nonspecific vaccine from 86 to 89 for the younger age group and from 1283 to 1335 for the older age group. Decreasing $\beta$ by 10 percent decreased the deaths in the simulation without the nonspecific vaccine in the younger group from 86 to 82 and from 1283 to 1212 in the older age group. 





\section{Conclusion}

In this paper we have examined some similar concepts in the renormalization group equations of high energy physics and pandemic modeling based on differential equations. This include the nonlinear behavior of the differential equations, the dependence on the initial conditions and renormlalization group evolution, the similar nature of the contact matrix between compartment groups in epidemic modeling and the notion of operator mixing and fixed points in the renormalization group equations. We put these features together in  models with mixing between different age groups, mixing with different countries and the effect of waning of nonspecific vaccines that were designed for different diseases but can lessen the severity of infection and mortality for a  pathogen associated with a pandemic. In the future it would be interesting to extend these case studies using more data with respect to different regions, age demographics and age related mortality. It will also be interesting to extend the contact matrix study with different countries as travel is resumed between countries with lower rates of infection. Finally we would like to update the model associated with waning of nonspecific vaccines as well as  specific vaccines as more data becomes available from clinical trials about the effect of nonspecific and specific vaccines on the rate of mortality for the pandemic.  

\section*{Acknowledgements}
This manuscript has been authored by employees of Brookhaven Science Associates, LLC under Contract No. DE-SC0012704 with the U.S. Department of Energy.

\end{document}